# The Blockchain Revolution: Insights from Top-Management

Theodosis Mourouzis[1,2]   Chrysostomos Filipou[2]

[1] UCL Centre for Blockchain Technologies, London, UK

[2] Cyprus International Institute of Management, Nicosia, Cyprus

## Abstract

This is an exploration of Blockchain technology that is growing in popularity and it seems to be able to disrupt a plethora of industries. A research is being conducted to examine Blockchain's potential to be adopted by enterprises from different sectors as well as the parameters that could affect its adoption. Mostly known as the technology that underpins Bitcoin, this concept raised a significant interest within various markets. Blockchain offers a new approach to valued information management and sharing and it is introduced as a solution against the inefficiencies that affect the industry. Experts, infrastructure providers and banks can now work on this technology and explore its uses. This is a new technology journey with obstacles that will need to be overcome and it can not be clear yet what will eventually arise. Professionals from around the world express their views on the adoption of Blockchain by organisations and how these plan to support its deployment. Thoughts are shared in terms of the required budget and the parameters that can impact its adoption. There is a great interest in Blockchain technology and its revolutionary potential to modernize the world economy and this is only the beginning.

# Introduction

Blockchain is certainly another disruptive product of the digital economy. Quite often it is considered as a synonym of Bitcoin, however Blockchain is neither Bitcoin nor any other cryptocurrency, it is the underlying technology. More specifically, in the case of a virtual currency, Blockchain is utilised as a digital public ledger of all transactions related to the virtual currency and this is only one of its many possible applications.

Today, the technology is generating intense interest as it can be utilized in plethora of industries such as banking, finance, insurance, healthcare, government and professional services. However, Blockchain is still a nascent technology and it comes with change. And when something is new and brings change can be challenging. The value of this emerging and growing technology still needs to be clarified since there is a lack of awareness and understanding of Blockchain and how it works, especially in sectors other than banking

As a result, we conducted a survey that its primary purpose is to collect information and gain the perspective of senior executives and IT Decision-makers, from around the globe, in regards to the adoption of Blockchain technology. More specifically, the survey aims to gain an understanding of whether different organisations are either seriously considering adopting Blockchain or have already started utilising it and which factors could have a positive or a negative impact on their decision in terms of the adoption of Blockchain.

The sample in this survey is targeted to senior professionals, more particularly senior executives and IT Decision-makers from around the world. Ultimately, participants could provide data in regard to the perception of the market about Blockchain technology and share information so that we understand to what extent Blockchain is going to be adopted and which parameters senior professionals rely their decisions on.

# What is Blockchain?

According to L. (2015) the innovation of Blockchain technology is credited to a person or group of people under the name Satoshi Nakamoto. Although this name is connected to the creation of Blockchain, an undeniably ingenious invention, not much is known about its real identity.

Blockchain became popular as the heart of Bitcoin and other virtual currencies however, it seems to be more than this, especially, when it comes to its potential to transform and to modernize the global economy. Blockchain seems to be a new, a more sophisticated way of managing valued information. The way the particular technology evolves, at this initial stage of its existence, as well as the potential impact that could have on many domains of the society, jog our memory back to 1989 and recall the discussion around the invention of Tim Berners-Lee, the World Wide Web. More specifically, Don and Alex (2017), who are considered to be thought leaders of Blockchain technology, refer to Blockchain as the technology that will lead internet in a new era.

Undoubtedly, there are many similarities related to the magnitude of the impact that Blockchain can make. This triggered our curiosity to learn more about Blockchain as well as the interest to study and try to gain an understanding in relation to the perception of the market about this nascent technology.

According to Tapscott (2017),

> *"The Blockchain is an incorruptible digital ledger of economic transactions that can be programmed to record not just financial transactions but virtually everything of value."*

This is a concise definition that can also help in understanding the need, which has led to the development of Blockchain technology. Namely, the need of recording valued information electronically in a way that will enable the option, for anyone, to participate in the global economy through its personal computer conducting transactions without the need of a middle-man.

Blockchain is certainly another disruptive product of the digital economy. Quite often it is considered as a synonym of Bitcoin, however Blockchain is neither Bitcoin nor any other cryptocurrency, it is the underlying technology. More specifically, in the case of a virtual currency, Blockchain is utilised as a digital public ledger of all transactions related to the virtual currency and this is only one of its many possible applications. Swan (2015), in an effort to group those sectors where Blockchain technology can be applied, he divided the existing and its potential application into three categories. Currency and Contracts are Blockchain 1.0 and Blockchain 2.0 respectively, in addition to Blockchain 3.0, which is the category that finances applications such as those used in government, health care, science, literacy, culture and art.

From a technical perspective, Blockchain technology is also known as a distributed database that is different from the relational databases that currently dominate the market. The work of Schneider et al. (2016), indicates that a distributed database improves security, transparency and efficiency in contrast to the current relational database management model that the support of transactions and computations is conducted by databases that are used as central repositories. Transparency, security and efficiency are being improved by the control of a distributed database that does not rest with its owner like it happens with the relational database model.

More specifically, it is an ecosystem consisted of technology innovations such as encryption, mutual consensus verification and smart contracts that all together are responsible for organising and sharing data so that we enjoy better security, transparency and efficiency.

*Security:* What makes Blockchain more secure than the traditional database management model is the heavy-duty encryption involving public and private key to maintain virtual security.

*Transparency:* Blockchain resides on the network and it is extended out of the boundaries of a single institution. As a result, not a single institution is responsible for a transaction audit or to keep records. Therefore, it is impossible for someone to hide a transaction and this secures that all transactions are more traceable than the current way of handling this info.

*Efficiency:* Blockchain technology allows us to conduct a transaction within seconds via a smartphone and no documentation is required to be signed between the related transaction parties.

## Technology Innovations that work with Blockchain

According to Van de Velde et al. (2016), Encryption, Mutual consensus verification and Smart Contracts are the technology innovations a Blockchain is consisted of. In this section, these innovations will be described.

*Encryption:* This is the component that has an immediate impact on the security. In essence, a sophisticated mathematical calculation known as "hash" takes place every time a transaction between two parties is in progress. This increases the security and provides the ability for anonymity of sensitive data. Practically, it allows end-users that could be the participants of a transaction, to selectively disclose information to other parties.

*Mutual consensus verification:* As previously mentioned, Blockchain is a distributed database that everyone can update without the need of a governing authority. This implies that something else should govern or control the update process of the information that is stored on the distributed database. As a result, the mutual consensus verification takes place via consensus protocols that validate, safeguard and preserve transactions.

*Smart Contracts:* According to Crosby et al. (2016), smart contracts are programmes that are responsible to enforce all the conditions of an agreement between two parties. In other words, a smart contract is a contract without documents, which ensures that the agreed conditions have been applied. Otherwise, any kind of transaction between different parties can not be completed. It is worth mentioning that although the "smart contract" was invented, by Nick Szabo in 1994, it has not been utilised until the time Blockchain, via virtual currencies, appeared.

**How Blockchain works?**

A good starting point on our effort to describe how Blockchain works should be the description of each of the five principles that underlie the technology, as these are stated by Murck (2017).

i. *Distributed ledger:* A transaction between two or more parties is recorded in a database and the related information as well as its transaction history is accessible by each party. There is no need for an intermediary entity that will control the transaction because of Blockchain's ability to verify the information related to each transaction and secure that not a single party has the control.

ii. *Peer-to-peer Transmission:* The existence of the distributed database leads to the ability for a direct communication (peer-to-peer) between the participants of a

  transaction instead through an intermediary node. In other words, everyone can reach and exchange information with all others directly.

iii. *Transparency with Pseudonymity:* By default, a transaction in a Blockchain ecosystem occurs between Blockchain unique, for each party, addresses. However, each party has the option to provide proof of its identity to others.

iv. *Irreversibility of Records:* As soon as a block is created then it can not be altered because it is connected to a block that consists of information of a previous transaction.

v. *Computational Logic:* The fact that Blockchain is a digital mean for transactions, implies that a computational logic can be programmed. As a result, rules can be applied through sophisticated algorithms.

Schneider et al. (2016), in an effort to describe how Blockchain works, they used the process of creating a block that is added to an existing chain. Following this approach and considering the description of each aforementioned principle, a high-level description is provided below.

As it has been mentioned, Blockchain is a distributed database, which means that it is replicated across different nodes in a network. Its dataset consists of the information that is required for a transaction between different participants. Each transaction's information is split into different blocks, namely a block is created for each transaction. If we consider a virtual currency transaction we can easily understand that each block contains information related to the seller and the buyer, as well as the price and the contract terms, including other related details.

After the initialisation of the transaction and before the completion, which is the creation of the block that will be added to the chain of previous transactions, validation is required. The validation of each block is conducted by the entire network through the encryption that has been described earlier. The successful completion of the validation step depends on the result of a hash calculation that should match other nodes so that the block will be added to the chain. The following figure, (Figure 1) shows the description of such a process between two parties.

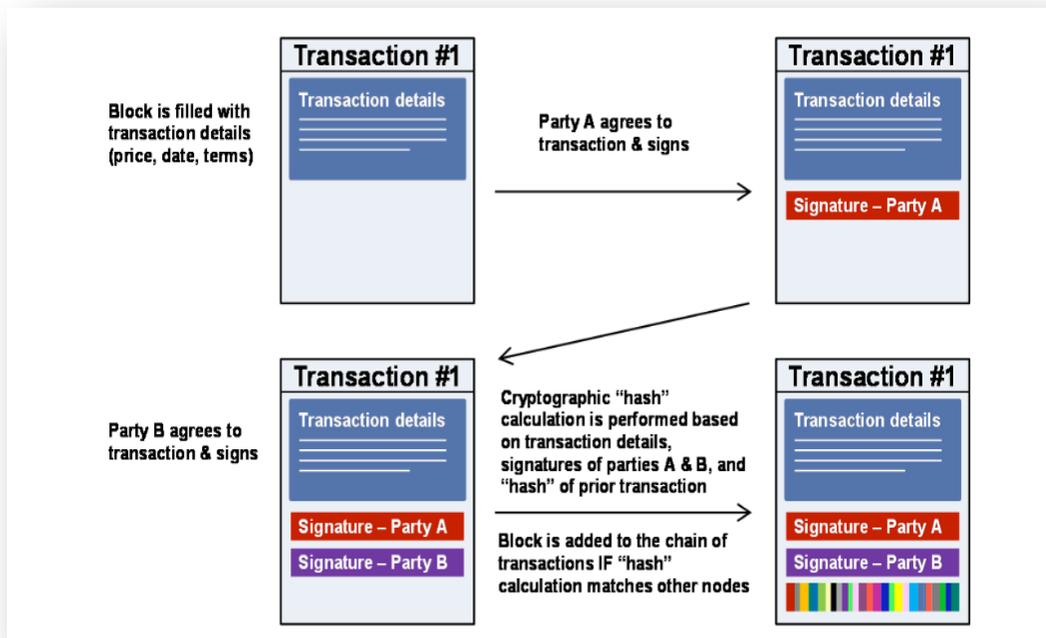

**Figure 1: Representation of the process followed for the creation and validation of a single block. (Schneider et al., 2016)**

The fact that the Blockchain ledger is distributed across a different location implies that the same information, which a block consists of, should be replicated to all different locations. Replicating multiple blocks across many nodes/locations can cost the alternation of a specific block (see location 5 on Figure 2). In such a case, the block will be corrected through mutual consensus verification.

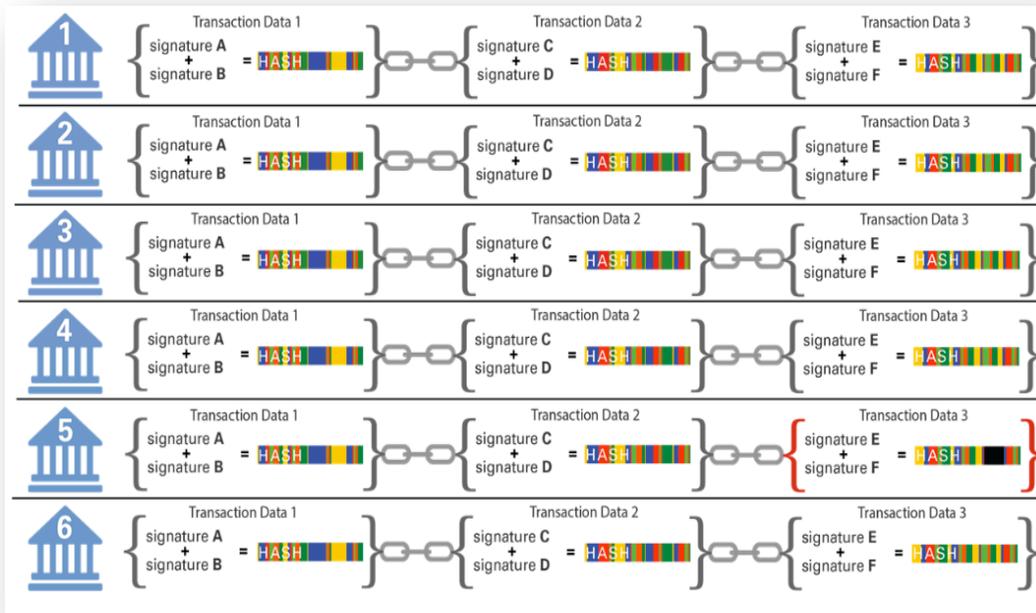

**Figure 2: Depiction of distributed ledger that is replicated across six different locations.(Schneider et al., 2016)**

## Profile of Survey's Participants

The target participants of the survey have been identified and contacted individually, mainly through social media. The pie chart below (Figure 3) confirms that the survey attracted people from around the world, (27.78%) United States of America, (52.79%) Europe, (5.56%) Canada, (2.78%) Africa, (2.78%) Israel, (2.78%) Japan and (2.78%) United Arab Emirates (UAE).

Figure 4 indicates the various sectors of the organizations that participate in this survey. The largest group of respondents with a percentage of 37.5% comes from the Financial Services sector that could be an indicator that Blockchain gains traction in this particular sector. Notably, the survey attracts responses from other sectors as well, like consulting, insurance, Information Communication Technology (ICT), government and fiduciary services.

The results of the question 2 and 3 shown on the Figure 5 indicate that the survey reached successfully the target audience with the characteristics that have been predefined. In other words, we managed to attract the decision makers from a broad range of small, medium-sized, and large organisations.

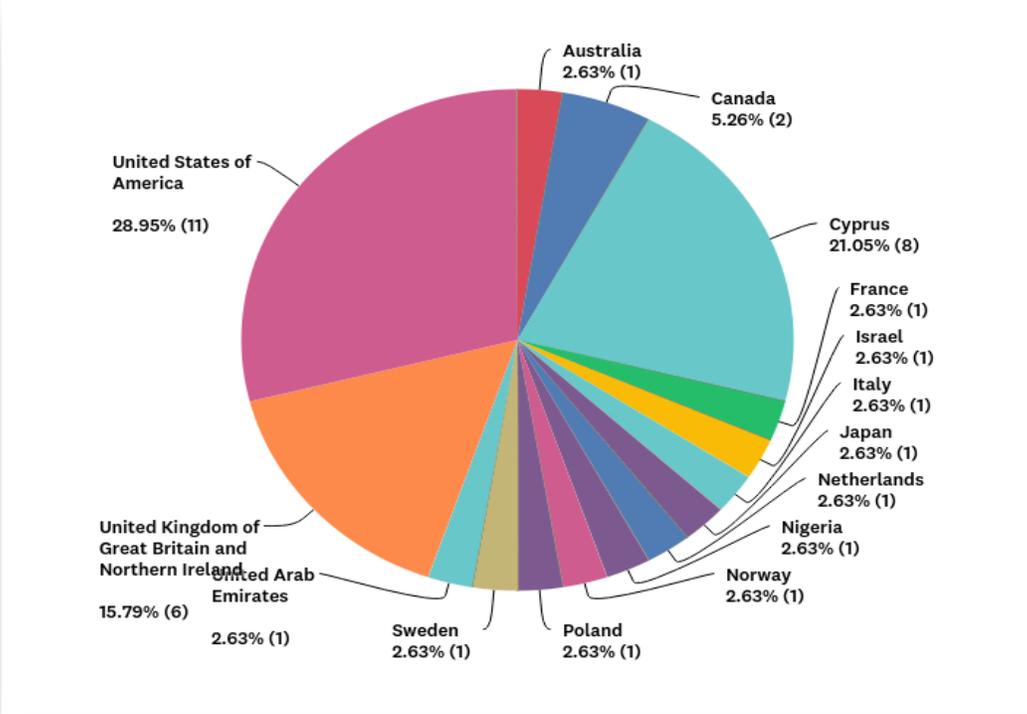

**Figure 3: Countries where survey participants work.**

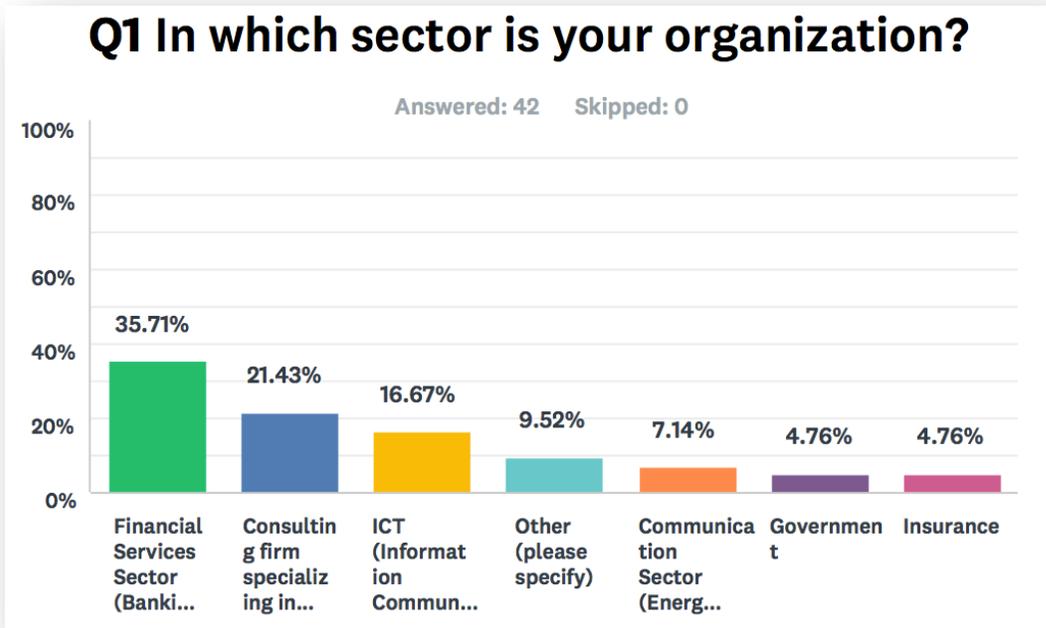

Figure 4: Sector of participants' organisations.

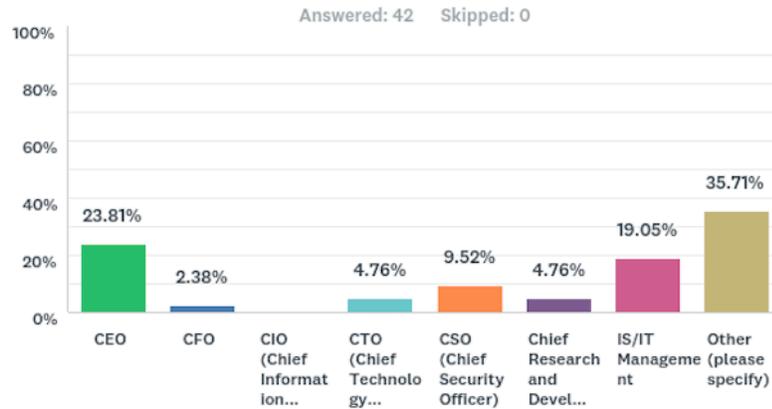

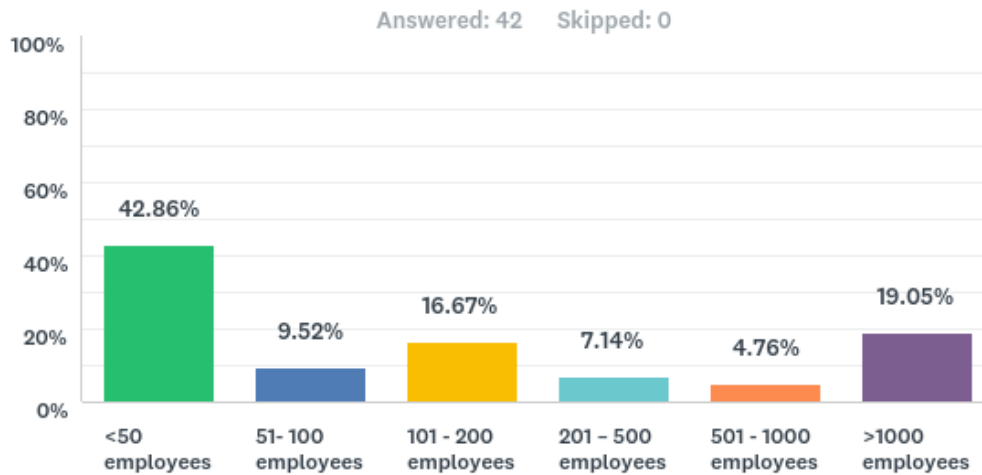

**Figure 5: Hierarchy level of survey participants and organisations' sizes.**

## To What Extent Organisations Adopt or Plan to Adopt Blockchain

As it is mentioned earlier, this survey aims to get an insight on the status of Blockchain in different organisations. The Figure 6 below illustrates a high percentage (52.38%) of organisations that explore Blockchain technology which is aligned to the extremely big percentage (73.81%) of the organisations that plan to invest in Blockchain in the next five years (Figure 7). The relatively high percentage (23.81%) of the organisations that are preparing for productive use can be an indicator of a market's extensive interest in Blockchain technology that will lead to a fast-pace adoption of the technology.

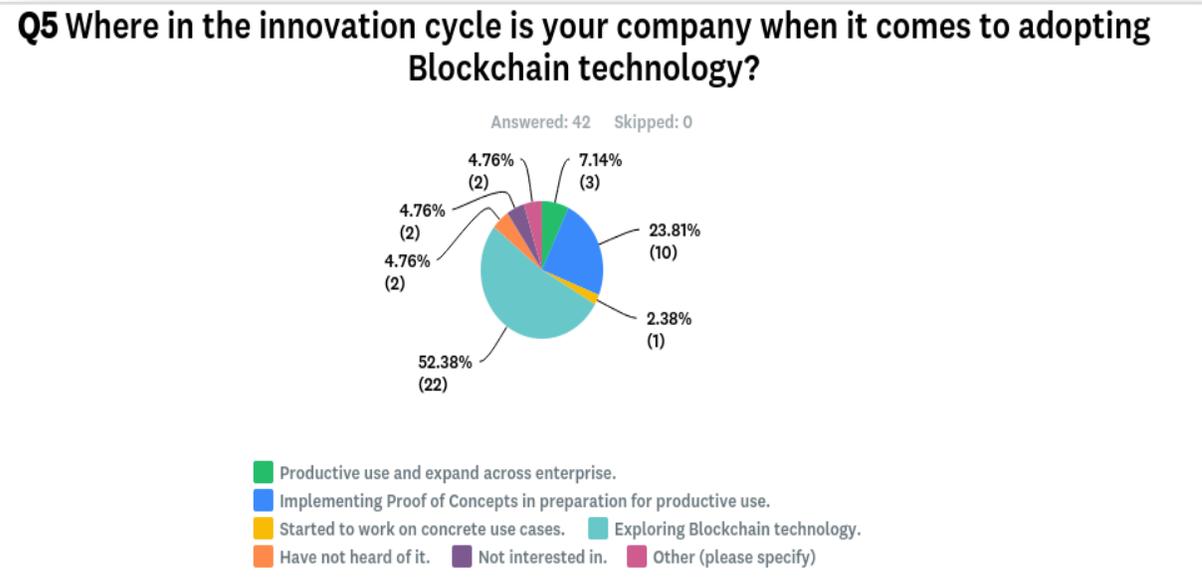

**Figure 6: Adoption status**

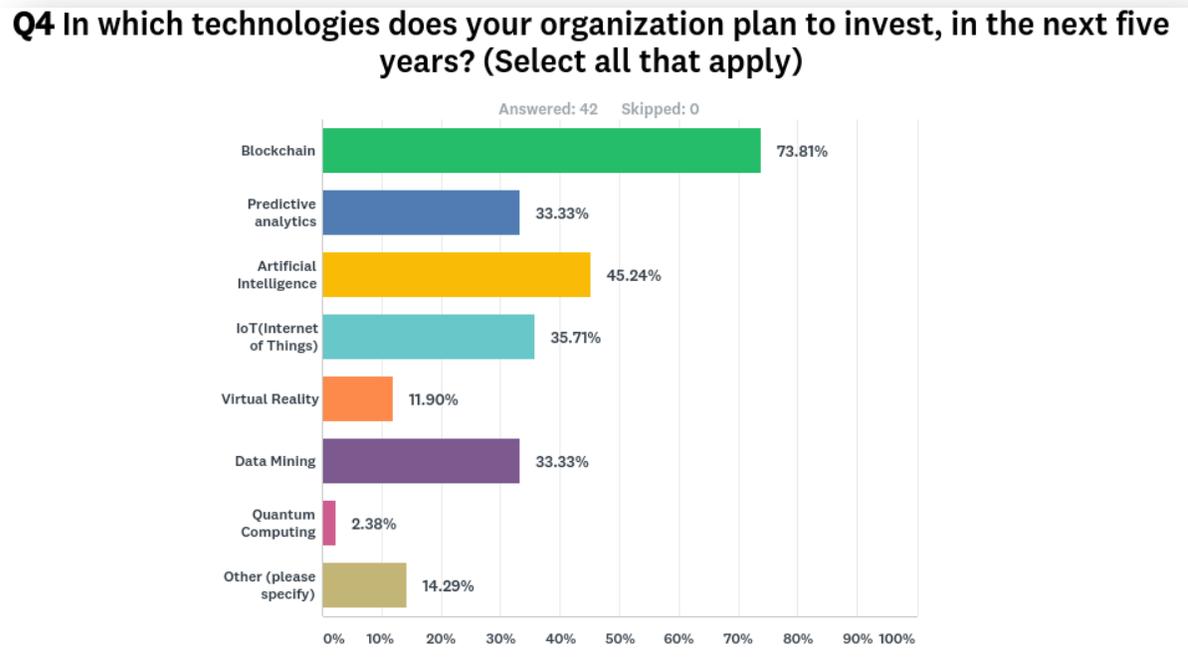

**Figure 7:** Attractive to organisations technologies

## How Organisations Plan to Support the Adoption of Blockchain

Figure 8 provides an insight on how organisations support or plan to support the deployment of Blockchain. Besides the fact that a big percentage (28.57%) of the survey participants are not aware about the plan of their organisation, a relatively high percentage (≈38%) of organisations seems to have already set up (26.19%) a dedicated Blockchain team or will do so (11.90%) in the next 12 months. It is interesting that 19% of the organisations participated in this survey plan to outsource the deployment of Blockchain solutions. In combination, with the results on Figure 9 that illustrates the importance (very important (8.96%), important (20.37%)) of expertise shortage on organisations decision to deploy Blockchain, we can understand why a significant percentage feels more confident to outsource the deployment of Blockchain instead of keeping it internal.

Figure 10 indicates that 21.43% of the participants replied that they have not yet decided which department will lead the efforts of Blockchain adoption. It is interesting that Business Strategy and R&D departments seems to be their first two options with a percentage of 19.05% respectively. An indicator that organisations may consider the adoption of Blockchain as a strategic decision.

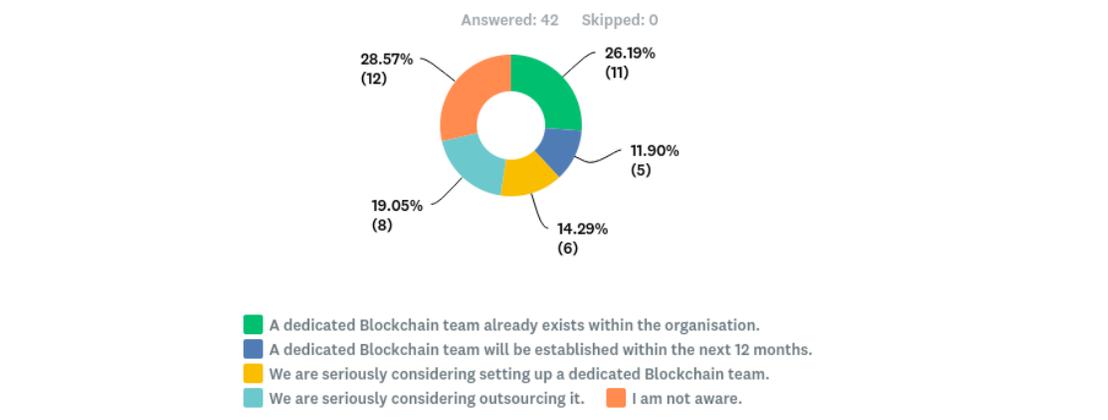

**Figure 8: Insight of how organisations plan to support Blockchain deployment.**

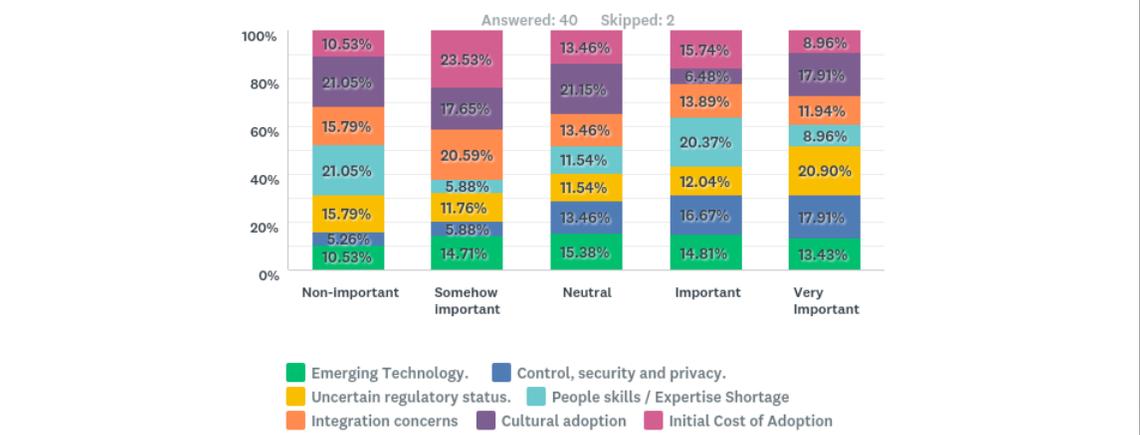

**Figure 9: Organisations perceptions in terms of Blockchain adoption challenges as those identified via the literature review.**

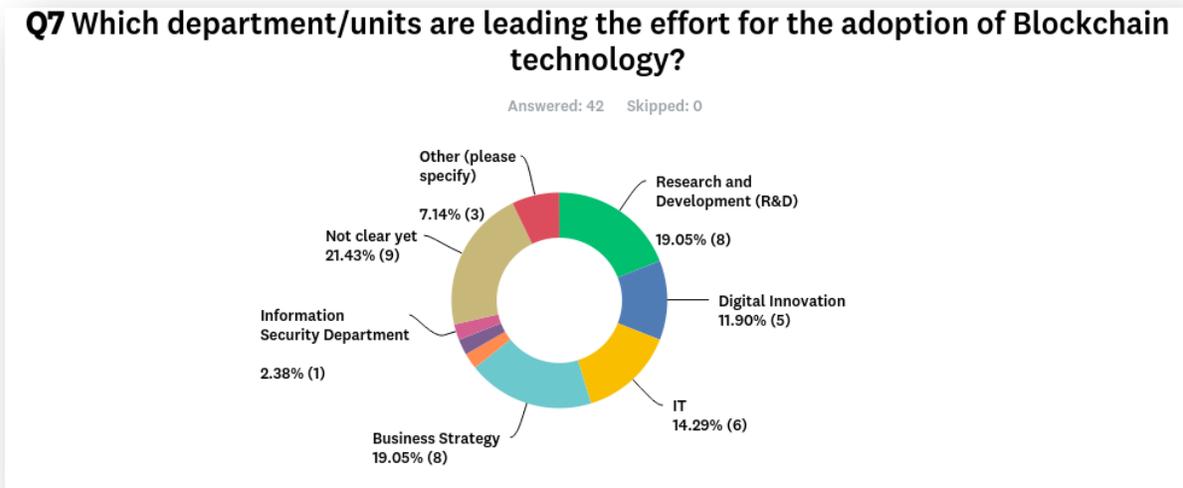

**Figure 10: The departments that will lead the effort for Blockchain adoption.**

## Budget

Budget is considered an important indicator of the pace a technology can enter the market of mainstream organisations. The results of our survey indicate that the vast majority of organisations do not have a budget however, they plan to create a budget linked to Blockchain in 2018 and afterwards. More specifically, 57.14% replied that their organisations do not have a budget while 33.33% have already created one (see Figure 11). The biggest percentage of those that have not created a budget yet, plan to do so in 2020 (28.57%) and 2021 (28.57%) (see Figure 12) that is justified from the results shown on Figure 14 illustrating that 72.5% of the surveyed participants anticipate that Blockchain will disrupt their business domain from 2019 to 2025.

In terms of the size of the budget that organisations are willing to place on Blockchain initiatives we are not allowed to be confident for the results of our research because of two reasons. Firstly, the sensitive nature of the information. Secondly, the number of organisations that have already linked a budget to Blockchain is relatively low (14 organisations). However, Figure 13 shows that 42.86% of the responses of the surveyed organisations that start investing in Blockchain, they currently have a budget that is bigger than 500,000 USD. A relatively big amount that could indicate the accelerated adoption pace of Blockchain.

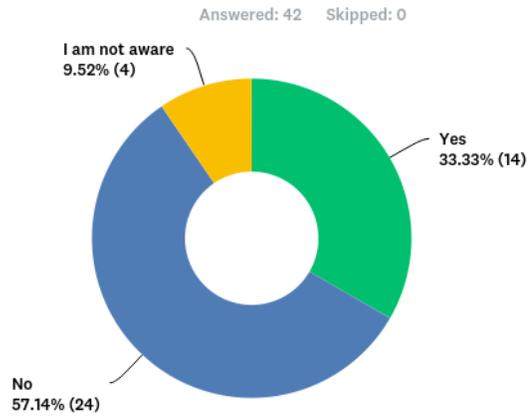

**Figure 11: Percentage of organisations that have Blockchain Budget**

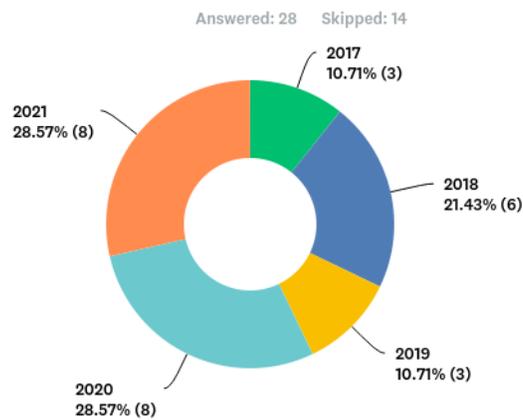

**Figure 12: When organisations plan to create a Budget linked to Blockchain.**

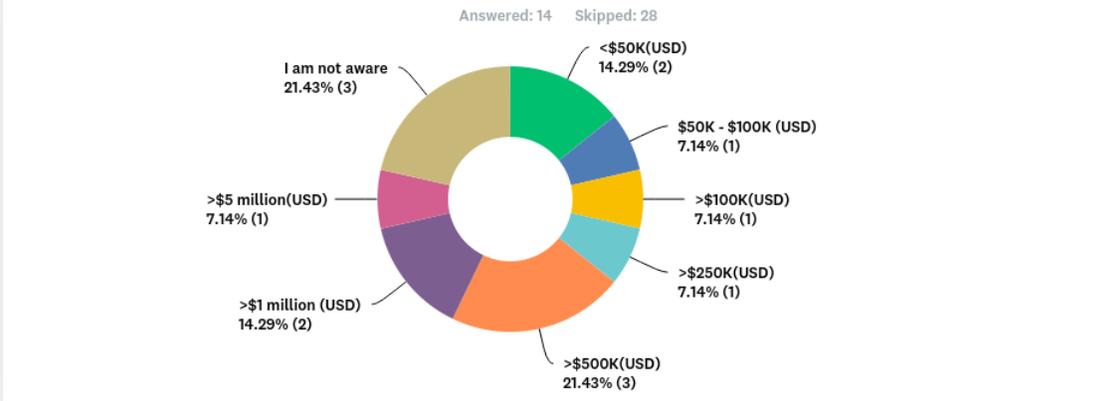

**Figure 13: The budget of organisations that have already created a Budget linked to Blockchain.**

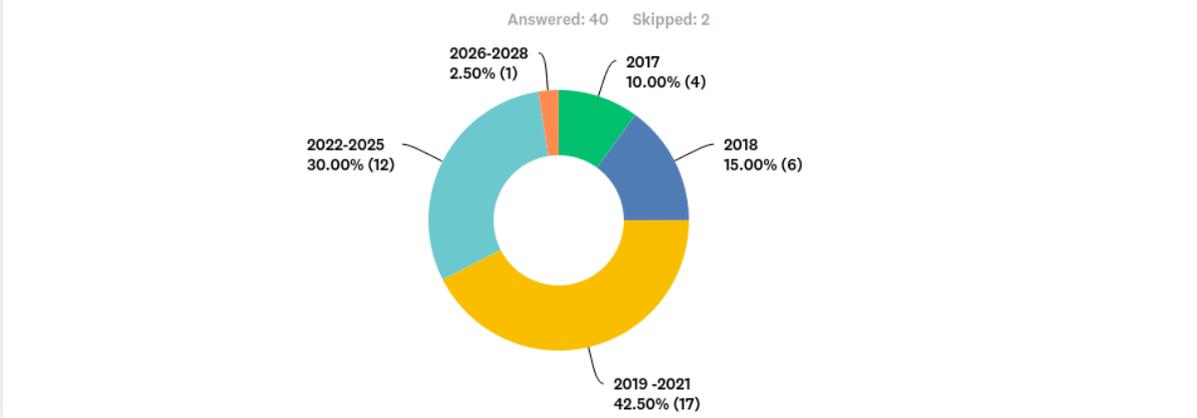

**Figure 14: When surveyed participants anticipate Blockchain to make the difference in their organisations and markets.**

**Parameters That Could Positively or Negatively Impact Blockchain adoption**.

The results of the questions 12, 13, 14 and 21 provide an insight on those factors that could impact the decision making related to Blockchain adoption.

Figure 15 indicates that surveyed participants consider the benefit of high quality data and the transaction transparency and immutability as the most important ones in relation to their organisations future plan with weighted average 4.35 and 4.17 respectively. The results could be a product of nowadays' trends related to the value of big data and the need for transparency in transactions. The latest is mostly related to financial services organisations that are obliged to comply with regulations and directives related to KYC/AML.

Figure 16 highlights those three sources that organisations mainly rely on so that they learn how and whether Blockchain can benefit them. More specifically, early adopters, thought leaders and market analysts with 70%, 55% and 42.5%, respectively are these main three sources.

The perception of surveyed organisations about the potential of Blockchain to disrupt their business domain as well as the challenges as those are defined by the available literature, can also impact organisations decision for Blockchain adoption. Figure 17 shows that 50% of the surveyed participants consider Blockchain capable enough to disrupt their organisation's sector while Figure 18 depicts the main concerns that would negatively impact the adoption rate of Blockchain. More specifically, 42.5% and 40% of the surveyed sample worries about the scalability and support & available documentation respectively. Although based on today's literature Blockchain seems to be a very reliable and secure solution, security with 35% is still among the first three factors that worries organisations the most. As a result, a rumour or a report for a security incident could easily negatively impact the adoption pace of Blockchain and affect its reputation.

It is interesting though that 65% (Figure 19) of the surveyed organisations are willing to immediately proceed with the adoption of Blockchain, if they are convinced that such a move will strengthen their value proposition. A high percentage (62.5%) of organisations are also willing to immediately adopt the technology as they see it as an opportunity to explore the technology and conduct an assessment as far as the potential for their organisation is concerned.

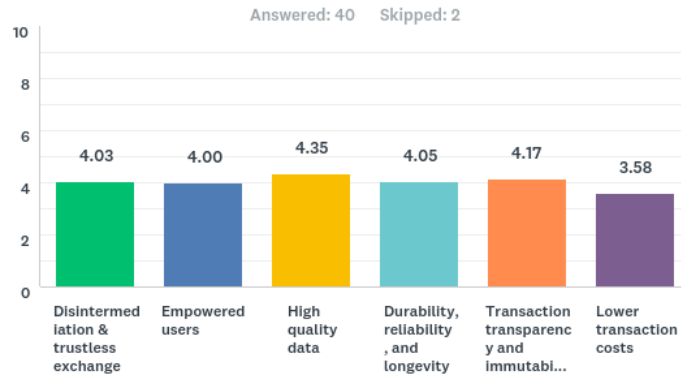

Figure 15: Importance of each benefit as those identified from the literature review.

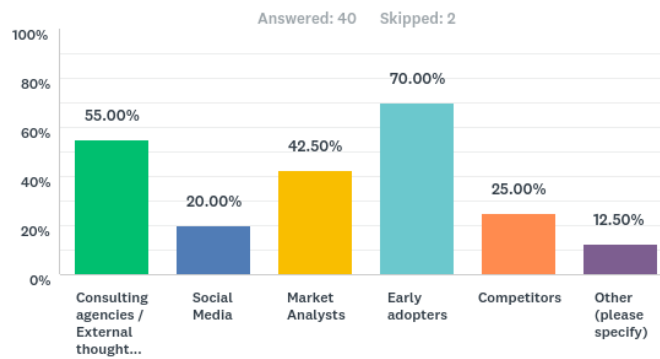

Figure 16: Sources that organisations rely on so that they learn about Blockchain.

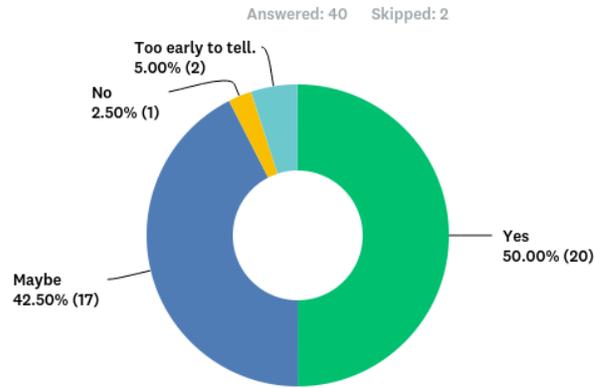

**Figure 17: Perception of surveyed organisations related to the potential of Blockchain.**

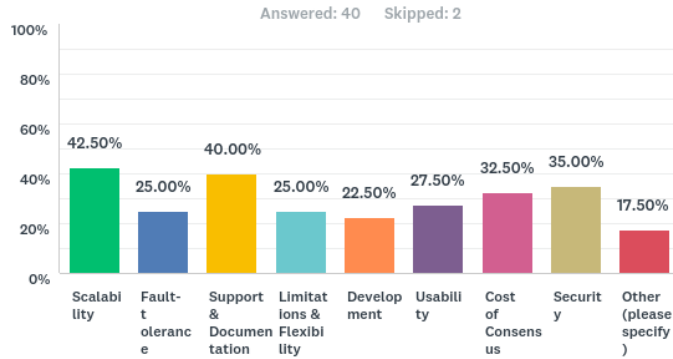

**Figure 18: Challenges**

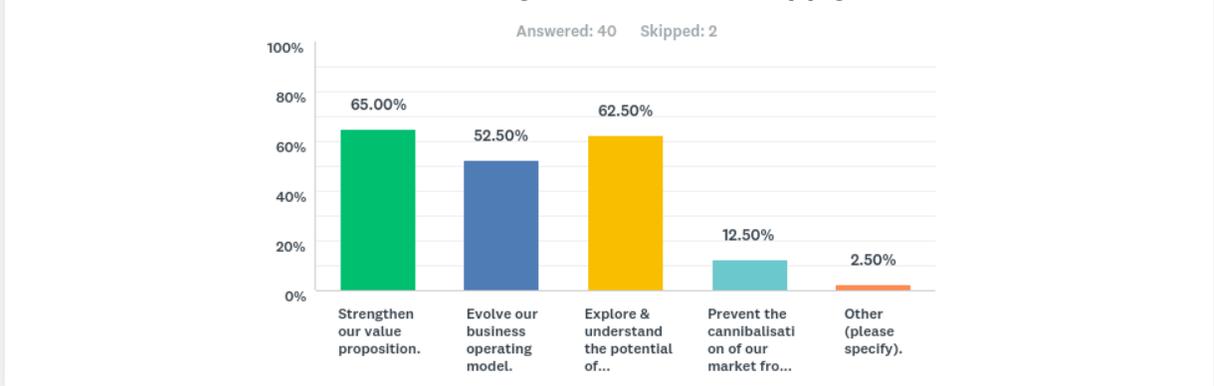

**Figure 19: What would lead to an immediate adoption of Blockchain.**

## Sectors where Blockchain could be adopted faster

According to the weighted average results that are shown on Figure 20 surveyed participants consider the sectors of professional services that will have significant impact. In particular, Banking with 4.78 and Finance and Accounting with 4.72 weighted average are on the top followed by the domain of Trade. The conclusion from the results of the question 16 can be confirmed by looking at Figure 21 where we can observe that all the related to those industries use-cases have the biggest weighted average score. For instance, contract management and regulatory compliance are among the most popular with 4.36 and 4.11 respectively.

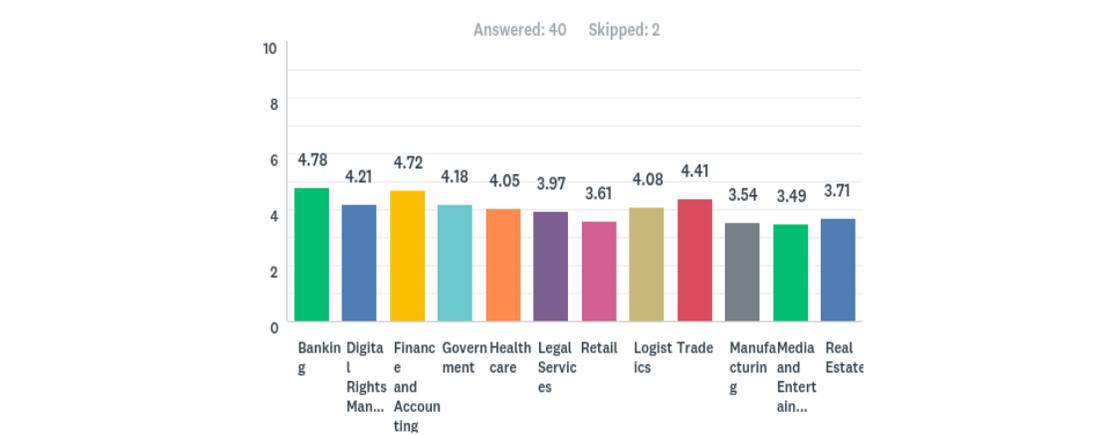

**Figure 20: Main Industries/Business Domains that could be significantly impacted**

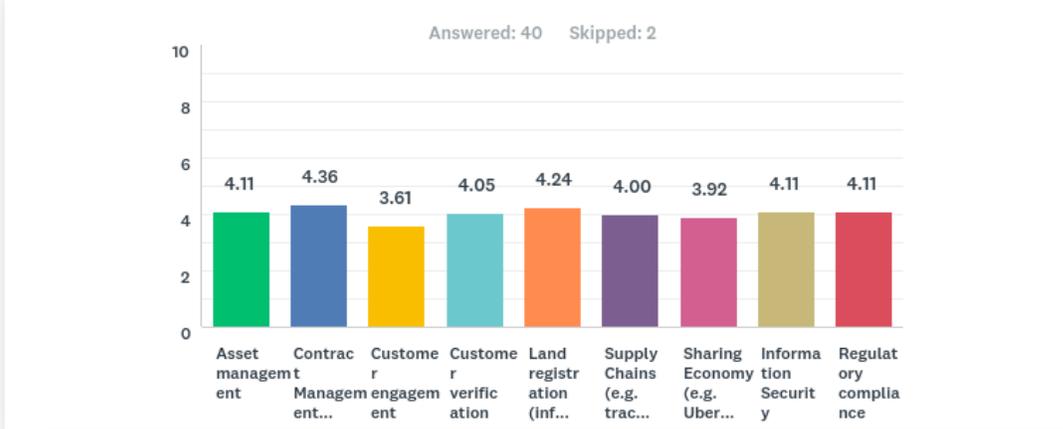

**Figure 21: Use Cases**

## Level of familiarity with the most popular platforms

The question 22 indicates that the level of familiarity of surveyed participants with the most popular platforms. With 35% of the participants to state that are very familiar and 35% to be familiar, with the platform called Ethereum (Figure 22). This can be an indicator of the correctness of the results of M Macdonald et al. (2016), research that has been presented earlier on this report as part of literature review. Also, this can confirm the competitive advantage of Ethereum.

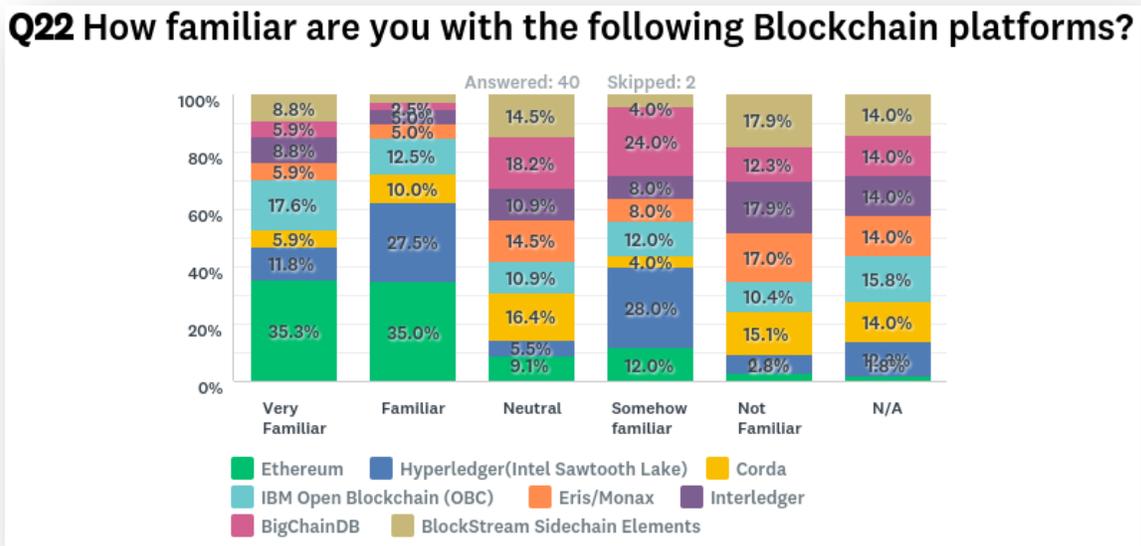

Figure 22: Familiarity with the most popular platforms

## The perception of surveyed participants regarding the future of Blockchain

The last two optional, open-ended questions, provided us with some interesting statements that depict surveyed participants' perception on why Blockchain will or will not finally disrupt the global economy. Question 23 and 24 were answered by 17 and 18 surveyed participants, respectively. Patterns among the responses were identified before the discussion of the results of the particular questions.

The following responses of the question 23, ("*Why Blockchain has the potential to revolutionize the world economy?*") linked Blockchain's success to disrupt the world economy with its ability to introduce better transparency, security and efficiency. It is worth mentioning that this is something that was previously mentioned as part of the literature review.

***Response 1:*** *"Through speed, relatively low cost and disintermediation".*

***Response 2:*** *"Secure".*

***Response 3:*** *"Shared ledgers substantially reduce cost and time in transaction processing. Decentralization allows quick creation of new networks."*

***Response 4:*** *"Transfer of ownership without a third party, communication mechanism to move data between enterprise systems".*

***Response 5:*** *"Stable, incorruptible currency".*

***Response 6:*** *"Can remove friction and delays in operations".*

The following response grabbed our attention as it is aligned with the literature review where it is already discussed that Blockchain success is linked with its versatility.

***Response 67*** *"Scalability and infinite adoption of use cases".*

As far the responses of the question 24 are concerned, it is interesting that many of the survey participants can not find any reason why Blockchain could fail to disrupt or revolutionize global economy. Indicative responses are quoted bellow.

***Response 1:*** *"hard to think of an example..."*

***Response 2:*** *"n/a"*

*Response 3:* "none"

*Response 4:* "it has"

*Response 5:* "It will revolutionise the world economy".

*Response 6:* "I believe regulators, central banks, and banks will try but will ultimately fail".

However, according to the following responses some of the surveyed individuals consider that regulations as well as the lack of understanding on behalf of governments and authorities will negatively impact or undermine Blockchain technology.

*Response 1:* "Government regulations, too many hands in the cookie jar and they make lots of money".

*Response 2:* "I believe regulators, central banks, and banks will try but will ultimately fail".

*Response 3:* "Scalability - no regulations - lack of mass understanding''.

*Response 4:* "Low level of understanding from public, government and regulatory authorities".

Herein, the data collected from the questionnaires were presented, discussed and have drawn interesting conclusions. Blockchain technology is handled by decisions makers worldwide who work in different sectors, in smaller or bigger companies with less than 50 or more than a thousand employees. The two biggest worries of the decision makers regarding this technology seem to be the lack of regulations and the shortage of expertise in the field, but this will not be a show-stopper if organisations are convinced that this technology can strengthen their value proposition. Also, the sector of professional services seems to show the greatest interest in Blockchain technology which implies a promising future for it. It is also worth mentioning that surveyed participants found it difficult to state an important reason why Blockchain can fail. In contrast, they enthusiastically responded to the question that asks their opinion on why this technology has the potential to revolutionise the world economy. To sum up, data shows that the future of Blockchain seems promising and revolutionary for the world economy and something that will be headlined a lot and gain traction in the near future.